\documentstyle[prl,twocolumn,aps]{revtex}

\begin{document}
\bibliographystyle{unsrt}
\input{psfig}

\twocolumn[{
\title{
DNA - Nanoelectronics:
Realization of a Single Electron Tunneling Transistor and a Quantum Bit 
Element}
\author {E. Ben-Jacob, Z. Hermon and S. Caspi} 
\address{
School of Physics and Astronomy, Tel Aviv University,
69978 Tel Aviv, Israel.}
\maketitle
\date{\today}
\vspace*{-.4in}

\begin{abstract}
\widetext
\leftskip 54.8 pt
\rightskip 54.8 pt
 
Based on the understanding that chemical bonds can act as tunnel junctions
in the Coulomb blockade regime, and on the technical ability to coat a DNA
strand with metal, we suggest that DNA can be used to built logical devices.
We discuss two explicit examples: a Single Electron Tunneling Transistor
(SET) and a Quantum Bit Element. These devices would be literally in the 
nano-meter scale and would be able to operate at room temperature. In
addition they would be identical to each other, highly stable and would 
have a self assembly property.

\end{abstract}
}]
\narrowtext

The quest for smaller and faster logical devices has persisted since 
the invention of the classical transistor. A novel idea of using single 
organic molecules as electronic circuit components has been proposed back 
in 1974 \cite{Aviram_Radner}. However, the difficulty to connect a single 
molecule to external leads prevented experimental verification of this 
idea until recently, when molecular junctions (acting as quantum dots) 
\cite{C60_1}-\cite{Reed}, and a carbon nanotube field-effect transistor 
\cite{Tans_Dekker} have been fabricated. 

Here we propose a new approach to make logical devices from molecules,
which is based on our understanding \cite{DNA_Charge_Solitons} that the
phosphate bridges in DNA can act as tunnel junctions in the Coulomb 
blockade regime, and on the technical ability to coat a DNA strand (and 
other molecules) with metal, thus forming a conductive wire with self 
assembly property \cite{Uri_Erez}. Our understanding is supported by the
observations of Reed et al. \cite{Reed}, who demonstrated experimentally
that chemical bonds act as tunnel elements. Combining the above conceptual 
and technical developments, we suggest to utilize the chemical bonds in 
DNA (or other molecules) to build logical devices. These would be literally
in the nano-meter scale and would be able to operate at room temperature. 
The operation principle of the proposed devices is the single electron 
effect, which makes them extremely fast. Since they are made from specific 
molecules, the devices would be completely identical to each other. They 
would also be highly stable due to the stability of the chemical bonds. 
The devices would inherit the self assembly property, which can be used 
to create complicated networks consisting of many elements.

\begin{figure}
\centerline{
\hbox{\psfig{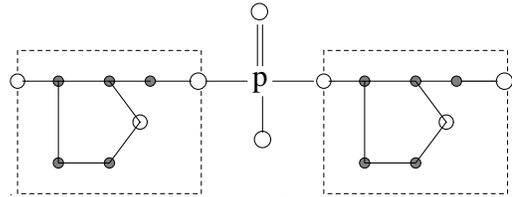}}}
\caption{A schematic image of two 'grains' in DNA connected by a P-bond.
The dark circles represent carbon atoms and the white circles oxygen atoms.}
\label{grainsP_fig}
\end{figure}

A DNA strand is made of units (or grains), composed of a sugar and a base. 
The grains are connected by phosphorus bridges (P-bonds), while 
complementary bases in different strands are connected by hydrogen bonds 
(H-bonds). We propose that a P-bond forms a tunnel junction for a net charge
\cite{DNA_Charge_Solitons}. (By 'net' charge we mean the deviation from the 
charge distribution of the unperturbed DNA.) The tunneling is either 
stochastic (like a normal tunnel junction) or coherent (like a mesoscopic 
Josephson junction), according to the coupling to the environmental degrees 
of freedom. The origin of this tunnel junction are the two oxygen atoms 
transversely connected to the phosphorus atom (see Fig.~\ref{grainsP_fig}). 
These oxygens share three electrons with the phosphorus, giving rise to two 
$\sigma$ bonds and one $\pi$ bond. As the $\pi$ electron can be shared with 
both oxygens, it resembles an electron in a double well potential and 
occupies the lowest level. When an additional electron approaches the well, 
it encounters a barrier due to the energy gap to the next level of the well. 
However, since this barrier is narrow and not very high, the approaching 
electron can tunnel through it.

\begin{figure}
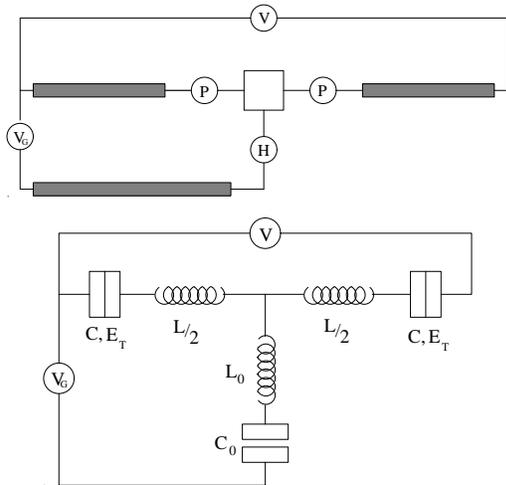

\centerline{
\hbox{\psfig{figure=schem.ps,height=1.0in}}}
      \vspace{0.1in}
\centerline{
\hbox{\psfig{figure=circuit.ps,height=1.4in}}}
\caption{A schematic image of a DNA SET transistor (above), and the 
equivalent electrical circuit (below). P denotes the P-bonds between the 
sugars, and H denotes the H-bond between the bases. $V$ and $V_G$ are the
external and the gate voltages, respectively. $C$ and $E_T$ are
the capacitance and tunneling energy of the P-bond, $C_0$ is the capacitance
of the H-bond, and $L$ and $L_0$ are the longitudinal and lateral 
inductances, respectively.}
\label{SET_fig}
\end{figure}

The H-bonds have a capacitive property. The proton in the H-bond can 
effectively screen a net charge density on either side of the bond by 
shifting its position towards this side. As a result, the net charge 
accumulates on the sides of the H-bond, and the bond can be viewed as a 
capacitor. The grains themselves have inductive properties, stemming from the
hopping of additional electrons. The notations are shown in Fig. 
\ref{SET_fig}.

According to the picture presented above, the DNA molecule inherently 
possesses all the properties needed for logical devices. The fabrication of
these devices can be done using available DNA manipulation techniques. As an
example we show how to build a classical SET transistor 
(see Fig. \ref{SET_fig}). One should start with two strands (a main strand 
and a gate strand), and connect the end base of the gate strand to a 
complementary base in the middle of the main strand. Both strands should be 
metal-coated, except the grain in the main strand which is connected to the 
gate strand, and its two adjacent P-bonds. The connective H-bond should be 
uncoated as well. To do this, the method presented in Ref.~\cite{Uri_Erez} 
has to be generalized to enable selective coating. We expect it to be 
feasible if artificially made strands are used, so that the coated and 
uncoated parts are composed of specific, yet different sequences of bases. 
Before the coating the DNA molecule should be in solution containing an 
enzyme which can bound only to the parts which should not be coated. After 
the coating the enzyme is released, and one obtains the desired result. The 
metallic coated ends of the main strand can be now connected to a voltage 
source ,$V$, and the end of the gate strand to another voltage source, $V_G$,
which acts as a gate voltage. 

This DNA-made device has the structure of a SET transistor \cite{CHET}, 
i.~e., a grain connected by two tunnel junctions to a voltage source, and 
biased by a capacitive coupling to a gate voltage. Usually, when discussing 
the SET transistor, one neglects the inductive properties of the grain. This 
neglect is equivalent to the claim that the relaxation time in the grain
is much shorter than the tunneling time. The values of $L$ and $L_0$ in the
DNA molecule are not known, but should be very small, probably in the pico 
Henry range. Thus the relaxation time is very short, and we can neglect the 
inductances here as well. We also assume that the coupling to the environment
is strong such that the tunneling is incoherent. In this case the voltages 
across the two junctions are
\begin{eqnarray}
\label{SET_Voltages}
&V_1&=V{C_2\over C_1+C_2}-{Ne+V_GC_0\over C_1+C_2} \ , \nonumber \\
&V_2&=V{C_1\over C_1+C_2}+{Ne+V_GC_0\over C_1+C_2} \ ,
\end{eqnarray}
where $N$ is the number of surplus electrons in the grain, and $e$ is the
electron charge. We added suffixes to distinguish between the two tunnel 
junctions. The instantaneous rate of tunneling from the right across the 
first junction, say is calculated from the single electron energy levels
\cite{TUN_RATES}:
\begin{eqnarray}
\label{Rates}
r_1(V_1)&=&\int_{-\infty}^{\infty}{2\pi\over\hbar}\left|T(E)\right|^2
D_{gr}(E-E_{gr})D_r(E-E_r) \nonumber \\
&& \times f(E-E_r)[1-f(E-E_{gr})]\,dE \ ,
\end{eqnarray}
where $T(E)$ is the tunneling matrix element for an electron in a state of 
energy $E$, $f(E)$ is the Fermi distribution function, $D_{gr}(E)$ and 
$D_r(E)$ are the density of states for the grain and the right electrodes, 
respectively, and similarly $E_{gr}$ and $E_r$ are their highest occupied 
energies. Their difference gives rise to a Coulomb blockade of tunneling into
the grain. The system obeys the Fermi distribution due to the large coupling
to the environment. The other tunneling rates: $r_2(V_2)$, $l_1(V_1)$, and 
$l_2(V_2)$ have similar expressions. The probability that there are $N$ 
electrons in the grain at time $t$ is governed by the master equation:
\begin{eqnarray}
\label{Master}
{\partial\rho\left(N,t\right)\over \partial t}&=&\left[r_1(N-1)+l_2(N-1)
\right]\rho(N-1,t)+ \nonumber \\
&&[l_1(N+1)+r_2(N+1)]\rho(N+1,t)- \nonumber \\
&&[r_1(N)+l_1(N)+r_2(N)+l_2(N)] \rho(N,t)\ ,
\end{eqnarray}
where the transition rates (\ref{Rates}) are expressed as functions of $N$ 
using (\ref{SET_Voltages}). The I-V characteristics of the SET are obtained
by solving Eq. (\ref{Master}) numerically using the appropriate initial and 
boundary conditions \cite{CHET}-\cite{Anal_CHET}. A typical I-V 
characteristic for constant density of energy states and identical junctions
in the low temperature limit has a voltage threshold. In order to operate as
an transistor, $V_G$ is varied around the threshold voltage. For well 
functioning transistor characteristics, the current raise above the threshold
value should be as steep as possible. This situation can be achieved if the 
tunneling rates (or $RC$ times) of the two junctions are different, or if 
there is a gap in the density of energy states of the grain. As the DNA 
molecule is not conductive it possesses a natural energy gap. The gap can be 
enhanced by using a larger section of DNA containing several grains instead 
of a single one. This happens since long DNA chains have non-linear effects, 
resulting in the tendency of charges to form solitons 
\cite{DNA_Charge_Solitons}. This method has also the advantage that it is 
technically easier to leave a larger section of DNA uncoated than a single 
grain. The tunneling rates in the DNA SET are the same, as the two P-bonds 
are identical. This situation can be amended by attaching a chemical group 
to one of the P-bonds, thus altering its properties. 

\begin{figure}
\centerline{
\hbox{\psfig{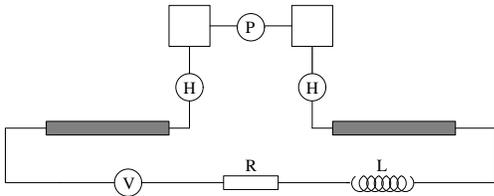}}}
\caption{A qubit made of one short DNA strand, attached to two long strands
by two H-bonds. The long strands are metal-coated and connected to an 
external voltage source, $V$, via resistance, $R$, and inductance, $L$.} 
\label{qubit_fig}
\end{figure}

As a second example for DNA-made logical device, we discuss a possible 
realization of a quantum bit (qubit), which is the fundamental element needed
for quantum computation \cite{Quant_Comp}. Several systems which can act as 
qubits have been recently proposed, included trapped ions \cite{Zoller} 
and Josephson junctions \cite{Alex_Gerd_Ziv}. The tunneling property of the
P-bond enables to use DNA to form a qubit realization similar to the one of
Shnirman et al. \cite{Alex_Gerd_Ziv}. Since the main concern in the operation
of qubits is maintaining quantum coherence over long periods of time, 
DNA-made devices can be used for quantum computation when the interaction with 
the environment is weak. In Fig. \ref{qubit_fig} we suggest how to build a 
qubit using three DNA strands: one short strand, containing two sugars and a 
P-bond in between, and two long metal-coated strands connected to the two 
sugars by H-bonds. This device has the same structure of a Josephson junction 
qubit \cite{Alex_Gerd_Ziv}, and should operate in the same manner. A detailed
study of a DNA-made qubit will be presented elsewhere.

To conclude, we have shown that the DNA molecule has the needed properties 
to make logical devices. We have discussed specifically how to build
a single electron tunneling transistor and a quantum bit element. Due to 
their nano-scale dimensions, conformity and availability, DNA-made logical
devices would have the advantage over the current solid state ones. In 
addition, the DNA-made devices would operate at room temperature and would
have a self assembly property.

This research is supported in part by a GIF grant G-0464-247.07/95.

\vskip 1cm

\end{document}